# Spin–orbit torques and magnetization switching in perpendicularly magnetized epitaxial Pd/Co$_2$FeAl/MgO structures

M.S. Gabor[1], T. Petrisor jr.[1,] M. Nasui[1], M.A. Nsibi[2], J. Nath[2], I.M. Miron[2]

[1]*Center for Superconductivity, Spintronics and Surface Science, Physics and Chemistry Department, Technical University of Cluj-Napoca, Str. Memorandumului, 400114 Cluj-Napoca, Romania*

[2] *SPINTEC, UMR-8191, CEA-INAC / CNRS / Université Grenoble Alpes, France*

We demonstrate efficient current induced spin orbit torque switching in perpendicularly magnetized epitaxial MgO(001)//Pd/Co$_2$FeAl/MgO heterostructures grown by magnetron sputtering. The advantage of such heterostructures for spin orbit torque MRAM devices is that they allow a unique combination of critical ingredients: i) low resistivity required for reduced electric energy consumption during writing, ii) high TMR ratio required for efficient reading and iii) strong perpendicular magnetic anisotropy for increased thermal stability.

Magnetization switching by current induced spin–orbit torques (SOTs) in heavy-metal (HM) /ferromagnet (FM) heterostructures with perpendicular magnetic anisotropy (PMA) is of great research interest for the development of electrically controlled spintronic devices[1-5]. One of the most important application of the SOTs is the three-terminal SOT-MRAM magnetic random-access memory. In this type of device, the information is written using current induced SOTs in a bottom HM/FM electrode, while the information is read using a top magnetic tunnel junction (MTJ) structure[1,3]. The HMs fall into two categories: high resistivity, such as Ta[2,3,5-10] or W[10-16], and low resistivity, such as Pt[1,4,8,10,17-19] or Pd[20,21]. While in-plane magnetized three-terminal devices have been fabricated using Pt-based structures[22,23], perpendicular magnetized three-terminal devices demonstrated so far are based on either Ta/CoFeB/MgO[3,24] or W/CoFeB/MgO[25]. The reason is that they usually grow without a preferred orientation, or even in an amorphous state. This allows the (001) crystallization at the CoFeB/MgO interface during subsequent annealing stages, which is essential to achieve a high magneto-resistive ratio[26] and a strong PMA[27]. However, the use of low resistivity HMs (Pt or Pd) is more desirable, because it reduces the Ohmic losses. The difficulty is that, when grown directly on Si/SiO$_2$ substrate or on a Ta buffer layer, the (111) fcc crystal face is energetically favorable and thus the (111) texturing is promoted.

This constitutes an important impediment, because the (111) orientation of the HM it is not compatible with the required (001) texturing for the upper FM layer.

Our approach is to use HM and FM materials which are (001) structural compatible, and grow (001) epitaxial HM/FM structures on a (001) MgO template. For the SOT layer we use Pd, which has already been shown to produce relatively large SOTs[21], and which, unlike Pt, can exhibit (001) growth on MgO. As the ferromagnet, we use Co$_2$FeAl (CFA), a Co-based full Heusler alloy[28] which has been shown to provide large magneto-resistive ratios in perpendicular magnetized MgO based magnetic tunnel junctions[29]. Therefore, in this paper we demonstrate that fully (001) epitaxial Pd/Co$_2$FeAl/MgO heterostructures showing perpendicular magnetic anisotropy can be grown on (001) oriented MgO substrates. We evidence relatively large SOTs and efficient SOT induced switching. Our results open an avenue for employing low resistivity HM in three-terminal SOT devices.

For our study, we used samples consisting of MgO (001)/Pd ($t_{Pd}$)/ CFA ($t_{CFA}$)/ MgO (1nm)/Ta (2nm). To obtain a complete analysis in terms of structural properties, SOTs amplitudes and magnetization switching we have varied the CFA thickness. Pd thickness is also varied in order to probe the thickness dependence of the SOTs. The samples were grown using a magnetron sputtering system having a base pressure lower than $2\times10^{-8}$ Torr. The metallic layers were grown by dc sputtering under an Ar pressure of 1 mTorr, while the MgO layer was grown by rf sputtering under an Ar pressure of 10 mTorr. The 2 nm thick Ta capping layer was deposited to protect the samples from oxidation due to atmospheric exposure. The structure of the samples was characterized by x-ray diffraction (XRD) experiments using a four-circle diffractometer. For magneto-transport measurements the stacks were processed by UV lithography and Ar ion milling into Hall bar devices with a channel size of $20\times10$ μm$^2$.

The XRD measurements were performed on relatively thick Pd (8 nm)/ CFA (10 nm)/ MgO (1nm) samples; they allow to verify that the growth is epitaxial. Since the epitaxial growth is initiated at the MgO/Pd interface, our observation that the thick samples are epitaxial implies that the thinner ones, which cannot be analyzed directly using conventional XRD, are also epitaxial. In order to quantify the spin orbit torques, we decrease the CFA thickness down to 2 nm. This thickness is ideal for determining the SOTs using the harmonic Hall techniques for in-plane magnetized materials. A further decrease of thickness is required for our last experiment for which we need perpendicular magnetic anisotropy. Here, we probe the SOT magnetization switching using magneto-optical Kerr effect.

Figure 1a shows *2θ/ω* x-ray diffraction patterns measured for the Pd (8 nm)/ CFA (10 nm)/ MgO (1nm) sample. Besides the peak corresponding to the MgO substrate, the patterns show only the (002) and

(004) CFA peaks and the (002) Pd one. This indicates an out-of-plane (00l) textured growth of the stack. Figure 1(b) shows φ-scan measurements for the CFA (022), Pd (022) and MgO (022) type reflections, recorded at a tilt angle ψ = 45°. The fourfold symmetry of the (022) CFA and (022) Pd reflections proves that the films are also in-plane oriented. Moreover, the φ = 45° angle separation between the CFA (022) and the MgO (022) and Pd (022) type reflections indicates the epitaxial relation MgO(001)[100] / Pd(001)[100] / CFA(001)[110], which imply an in-plane 45° rotation of the CFA crystal lattice relative to the Pd and the MgO one, as sketched in the inset of Fig. 1b.

In order to characterize the torques, we obtain in-plane magnetized samples by depositing stacks with a CFA layer thickness of 2 nm. We have also varied the thickness of the Pd layer from 1 to 10 nm. First, we evaluate the electrical resistivity of the samples. Figure 2(a) shows the electrical resistance of the structures as a function of the Pd layer thickness, measured by the four-probe technique in patterned samples. To extract the resistivity of the Pd layer we used a parallel resistor model in which we assumed that the resistance of the CFA layer remains constant and that the Ta capping layer is fully oxidized and does not contribute to the conduction. The inset of Fig. 2(a) depicts the Pd layer resistivity variation on the inverse Pd thickness. A roughly linear dependence is observed, that allows us to calculate a relatively low bulk Pd resistivity ($t_{Pd} \to \infty$) of $\rho_{Pd}^{\infty} = 12\ \mu\Omega$cm, in agreement with previous reports[21]. To quantify the anomalous Hall resistance $R_{AHE}$ and the effective anisotropy field $H_k$, we performed Hall measurements with the magnetic field $H_z$ applied out-of-plane. The $R_{AHE}$ shows a strong decrease with the thickness of the Pd layer, predominantly due to the current shunting through Pd, while the $H_k$ remains rather constant at about 1.1 T [insets of Fig. 2(b)].

The SOTs were evaluated using the harmonic Hall method developed by Avci *et al.*[30], which has the advantage of providing a straightforward method of excluding the thermo-electric effects. The measurement geometry is shown in Fig. 3(a). The measurement technique consists in injecting an AC current ($I_\omega = I \sin \omega t$) into the patterned stripe (along $\hat{x}$) and measuring the first ($R_\omega = V_\omega/I$) and the second ($R_{2\omega} = V_{2\omega}/I$) harmonic Hall resistances (along $\hat{y}$), while rotating the magnetization in-plane by applying an external in-plane rotating magnetic field ($H_{ext}$). Since the CFA films are in-plane magnetized, the $R_\omega$ provides information about the planar Hall effect as[30] $R_\omega = R_{PHE} \sin 2\varphi$, where $R_{PHE}$ is the planar Hall resistance and $\varphi$ is the azimuthal angle of the magnetization from the current direction [Fig. 4(a)]. Since the samples have a relatively weak in-plane uniaxial anisotropy, the $\varphi$ azimuthal angle of the magnetization is practically identical to the $\varphi_H$ azimuthal angle of the field [Fig. 3(a)]. The second harmonic Hall resistance $R_{2\omega}$ contains information about the SOT effective fields and it is given by[30]

$$R_{2\omega} = \left(\frac{1}{2}R_{AHE}\frac{H_{DL}}{H_{ext}+H_k} + R_{\nabla T}\right)\cos\varphi + R_{PHE}(2\cos^3\varphi - \cos\varphi)\frac{H_{FL}+H_{Oe}}{H_{ext}},$$

where $H_{DL}$ and $H_{FL}$ are the damping-like and field-like effective fields, $R_{\nabla T}$ is the second harmonic Hall resistance due to thermo-electric effects and $H_{Oe}$ is the Oersted field produced by the charge current. By fitting the $R_{2\omega}$ experimental data to the above equation, two contributions can be extracted: one which shows a $\cos\varphi$ dependence, and another which shows a $2\cos^3\varphi - \cos\varphi$ dependence [Fig. 3(c)]. The $H_{DL}$ and the $R_{\nabla T}$ are obtained from the slope and the intercept of the linear fit of the dependence of the $\cos\varphi$ contribution on the inverse of the sum of the external and anisotropy fields [Fig. 4(d)]. The sum $H_{FL} + H_{Oe}$ is determined from the slope of the linear fit of dependence of the $2\cos^3\varphi - \cos\varphi$ contribution on the inverse external field [Fig. 4(e)]. Finally, $H_{Oe}$ is calculated as $H_{Oe} = \mu_0 j_{Pd} t_{Pd}/2$, where $j_{Pd}$ is the charge current density through the Pd layer and $t_{Pd}$ is the thickness of the Pd layer, and it is subtracted from $H_{FL} + H_{Oe}$ sum to obtain $H_{FL}$.

Using the as determined values for the $H_{DL}$ and $H_{FL}$ we calculated the SOT efficiency per charge current density through the Pd layer and per unit electric field ($E$), defined by[31] $\xi^{DL(FL)}_{j_{Pd}(E)} = \frac{2e}{\hbar}\mu_0 M_s t_{CFA} \frac{H_{DL(FL)}}{j_{Pd}(E)}$, where $M_s$ is the saturation magnetization ($10^6$ A/m) and $t_{CFA}$ is the thickness of the CFA layer (2 nm). We have normalized the SOT effective fields to the electric field as well as to the charge current density through the Pd layer, in order to directly compare our results to literature data. The SOT efficiencies values are shown in Fig. 4 and are in agreement with previously reported ones for the (111) textured Pd/Co/AlO$_x$ system[21], indicating that CFA is a suitable ferromagnet for SOT switching.

Applications wise the PMA samples are the ones to be used due to their higher thermal stability. Therefore, we have probed SOT induced magnetization switching in structures showing PMA. To obtain the PMA we fabricated samples having a CFA layer thickness of 0.9 nm. Figure 5 shows representative hysteresis loops measured with the magnetic field applied perpendicular or parallel to the sample surface. The out-of-plane loop shows a squared shape with full remanence, indicating that the CFA films are out-of-plane magnetized. Moreover, the in-plane hysteresis loop shows a continuous rotation of the magnetization up to saturation, a comportment which is typical for a hard axis of magnetization. In order to extract the relevant anisotropy parameters, we have fitted the hard axis loop within the Stoner–Wohlfarth (SW) coherent rotation model[32]. We have considered the energy functional $E = (K_1 - 2\pi M_s^2)\sin^2\theta_M + K_2\sin^4\theta_M - M_s H\cos\theta_M$, where $K_1$ and $K_2$ are the first and second order magnetic anisotropy constants, $\theta_M$ is the magnetization polar angle measured from the perpendicular direction and the last term is the Zeeman energy. The experimental data were fitted by minimizing the total energy and using $K_1$ and $K_2$ as adjustable paraments. The results of the fit shown in Fig. 5(b) allowed us to determine $K_1 = (6.95 \pm 0.2) \times 10^6$ erg/cm$^3$ and $K_2 = (1.6 \pm 0.2) \times 10^6$ erg/cm$^3$. Using the as determined constant, we calculated the effective perpendicular anisotropy field $H_k^{\text{eff}} = 2K_1/M_s + 4K_2/M_s - 4\pi M_s$ to be around 0.75 T.

The samples are patterned into the same shape as for the torque measurements (asymmetric Hall crosses), only this time, in order to maximize the current density, the electric current pulses were applied along the narrower arm of the cross. A static in-plane magnetic field ($H_x$) was applied parallel to the electric current to allow bipolar SOT switching. The magnetization reversal was probed by wide field Kerr microscopy. A perpendicular magnetic field, larger than the coercive field of the device, was applied before injecting the electric pulses to ensure a fully saturated initial state. The image of the saturated state is used as a reference image, which is subtracted from images obtained after further applying current. The resulting differential images contain only the magnetic contrast of the reversed areas. Figure 6 (a) shows the differential Kerr images of magnetic switching for a series of 1000 current pulses of 2.14×10$^{12}$ A/m$^2$ having a width of 8 ns and with an in-plane magnetic field of 80mT. The dark (or bright) contrast indicates the regions where the magnetization has reversed. We observe that the switching takes place in the entire area of the narrower arm where the current density is high. The larger areas, where the current density is lower do not switch; their reversal would require larger current pulses. Furthermore, in Figure 6 (b) we plot the reversed area, versus the in-plane applied bias field. This indicates that the switching occurred progressively by nucleation and domain wall propagation (Figure 6c), which is typical for SOT switching[33,34]. Last but not least, we note that the switching current density in the Pd/CFA heterostructures is of the same order of magnitude as the current density required for switching the more conventional Pt/Co/AlO$_x$ tri-layers[4,19].

In summary, we investigated the spin-orbit torques in epitaxial Pd/CFA/MgO magnetic heterostructures grown by magnetron sputtering on (001) single crystal MgO substrates. The damping-like $\xi_E^{DL}$ ($\xi_{j_{Pd}}^{DL}$) and field-like $\xi_E^{FL}$ ($\xi_{j_{Pd}}^{FL}$) spin-orbit torque efficiencies per unit electric field (charge current density) show a monotonous increase with increasing the thickness of the Pd layer with no obvious saturation up to 10 nm. The relatively large SOT efficiency is confirmed by SOT switching of perpendicularly magnetized Pd/CFA/MgO heterostructures for current densities of the same order of magnitude as those reported for Pt/Co/AlO$_x$ tri-layers. Our results prove that the use of Heusler alloys can combine the energy efficiency of the SOT switching in low resistivity materials with the possibility of high TMR in perpendicular magnetized MgO based tunnel junctions, thus offering a platform for the development of energy efficient SOT memories.

M.S.G, T.P. and M.N. acknowledge the financial support for this work from MRI-CNCS/UEFISCDI through PN-III-P1-1.1-TE-2016-2131-SOTMEM grant number no. 24/02.05.2018. M.A.N, J.N. and I.M.M. acknowledge funding for this work from the European Research Council (ERC) under the European Union's Horizon 2020 research and innovation program (grant agreement No 638653 – Smart Design).

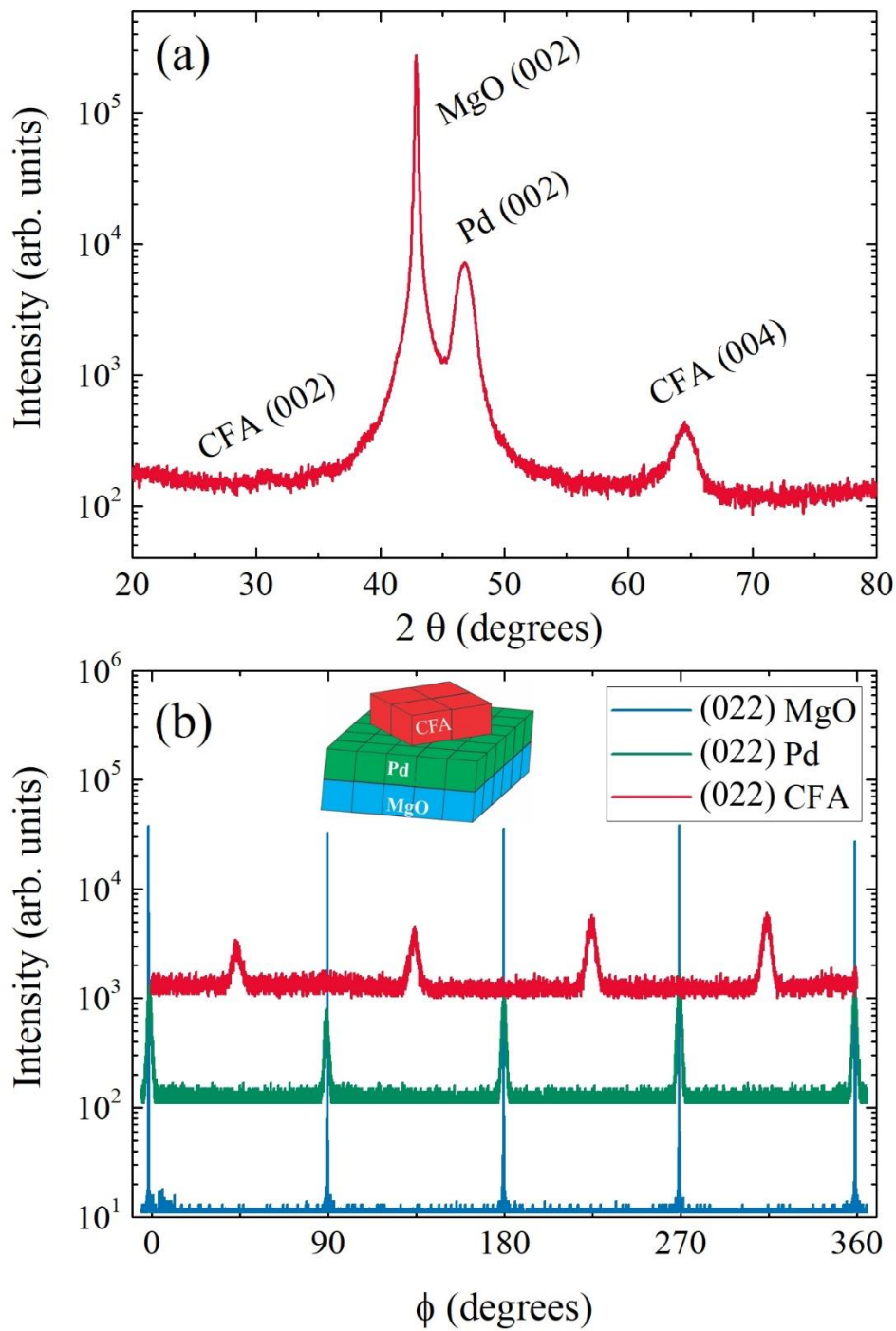

Fig. 1. (a) 2θ/ω x-ray diffraction pattern recorded for the Pd (8 nm)/ CFA 10 nm/ MgO (1nm) sample. (b) φ-scan measurements for the CFA (022), Pd (022) and MgO (022) type reflections. The inset shows a schematic of the CFA, Pd and MgO crystals lattice geometry.

Fig. 2. (a) Electrical resistance of the Pd ($t_{Pd}$)/ CFA 2 nm/ MgO (1nm) structures as a function of the Pd layer thickness. The inset depicts the Pd layer resistivity dependence on the inverse Pd layer thickness. (b) Typical anomalous Hall measurement performed with the magnetic field $H_z$ applied out-of-plane. The insets show the dependences of the anomalous Hall resistance $R_{AHE}$ and the effective anisotropy field $H_k$ on the Pd layer thickness.

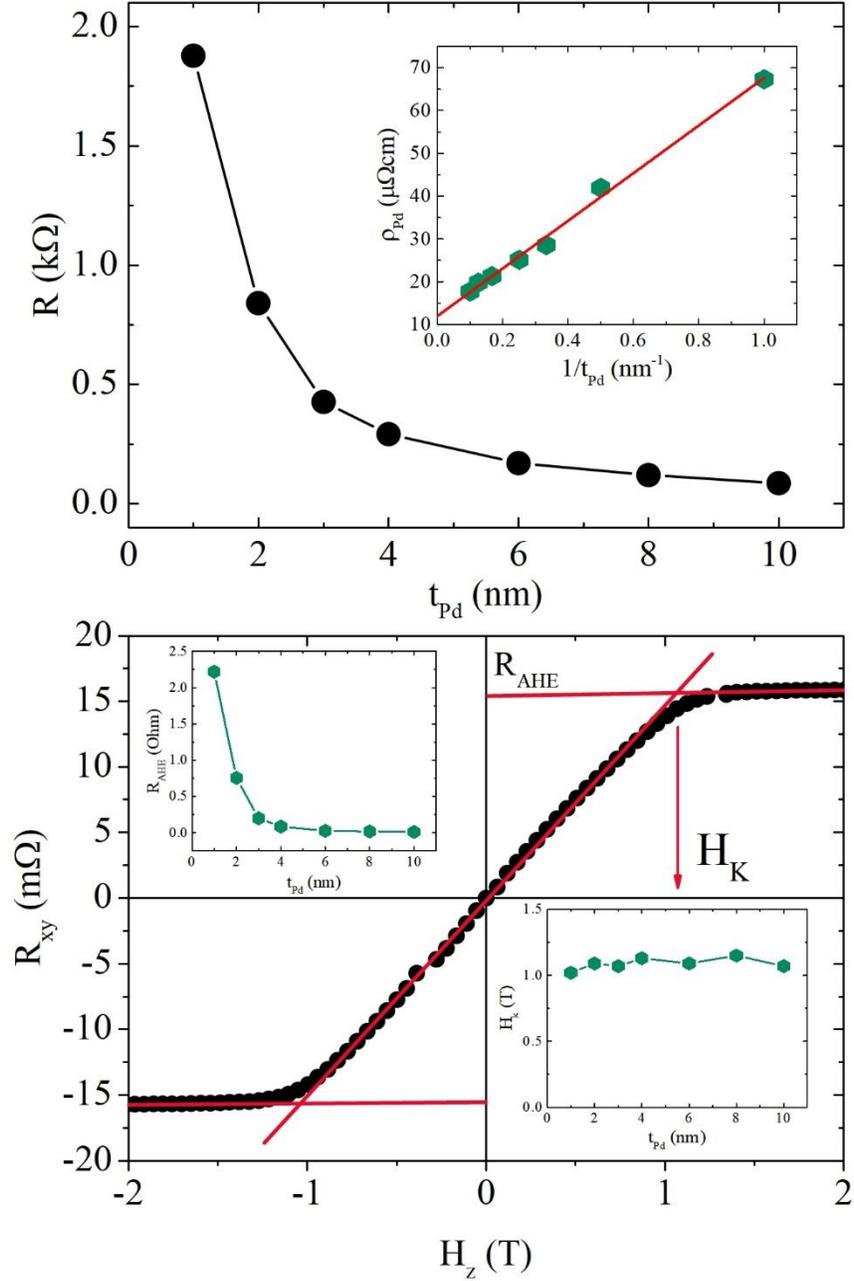

Fig. 3. (a) Schematic illustration of the Pd/CFA/MgO structure and harmonic Hall magneto-transport measurement geometry. (b) First and (c) second harmonic Hall resistances as a function of the azimuthal angle of the external in-plane magnetic field with the current direction, measured for a representative Pd (8 nm)/CFA (2 nm)/MgO (1 nm) sample. The points denote experimental data while the continuous lines are fits using the equations from the main text. Dependence of the second harmonic Hall resistance (d) $\cos\varphi$ – contribution on $1/(H_{ext}+H_k)$ and (e) $2\cos^3\varphi - \cos\varphi$ – contribution on $1/H_{ext}$, used to extract the $H_{DL}$ and the $H_{FL}$. The straight lines are linear fits to the data.

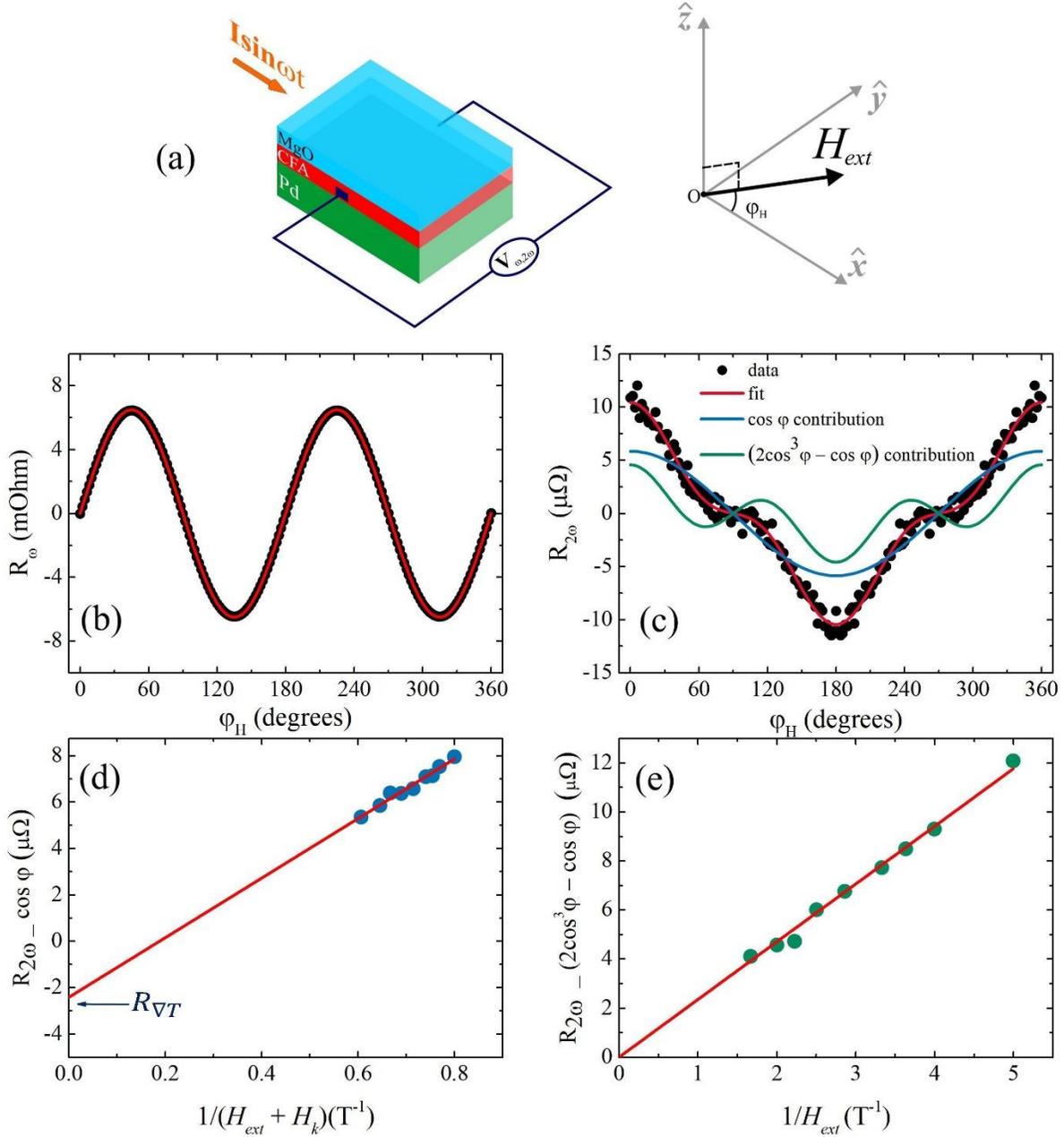

Fig. 4. The SOTs efficiencies per unit charge current density through the Pd layer ( $\xi^{DL}_{j_{Pd}}$ and $\xi^{FL}_{j_{Pd}}$ ) and per unit electric field ( $\xi^{DL}_{E}$ and $\xi^{FL}_{E}$ ) plotted versus the thickness of the Pd layer.

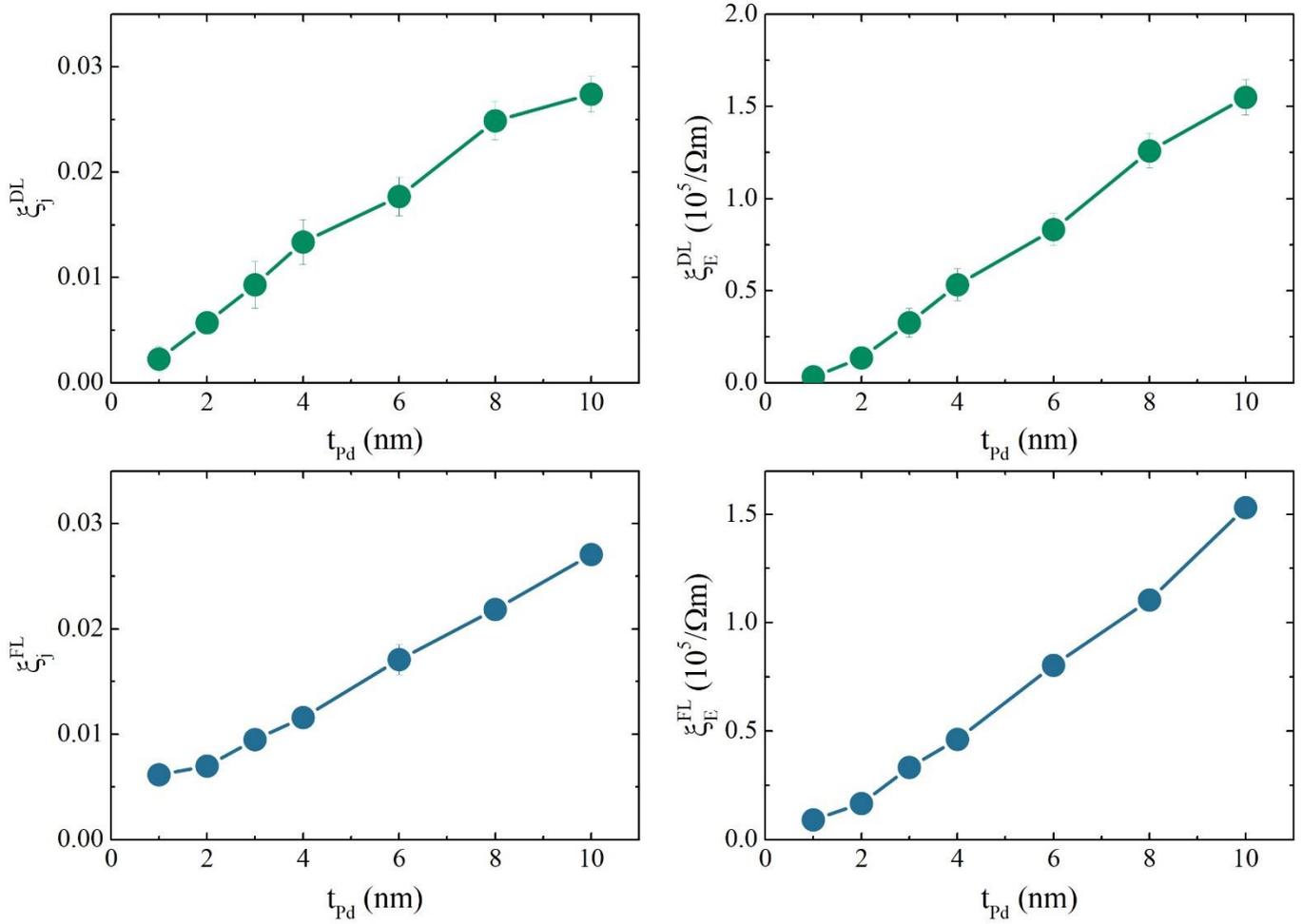

Fig. 5. (a) Perpendicular-to-plane and (b) in-plane hysteresis loops measured for the Pd (8 nm)/ CFA 0.9 nm/ MgO (1nm) sample, indicating the presence of perpendicular magnetic anisotropy. The points stand for experimental data while the red solid denotes the result of the fit within the SW model described in the text.

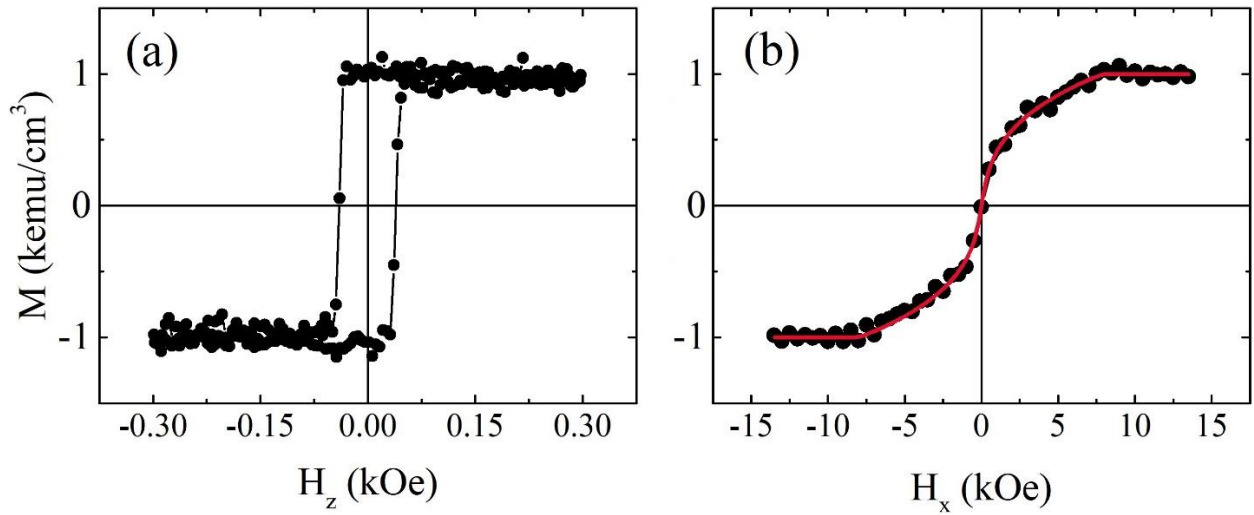

Fig. 6. (a) Differential Kerr microscopy images of the Hall cross magnetic switching by current pulses. The dark contrast in the top image indicates magnetization changing from the down state to the up state and the bright contrast in the bottom image indicates the reversed scenario. (b) The reversed area, imaged in (a), versus the in-plane applied bias field $H_x$ at a constant current density $J \approx 2 \times 10^{12}$ A/m². The electric current J flows from the right to the left in (a) and the direction of $H_x$ is opposite to that of the current. (c) Differential Kerr microscopy images of gradual switching. The nucleation occurs first on the edges of the device and the reversal is completed by domain wall propagation.

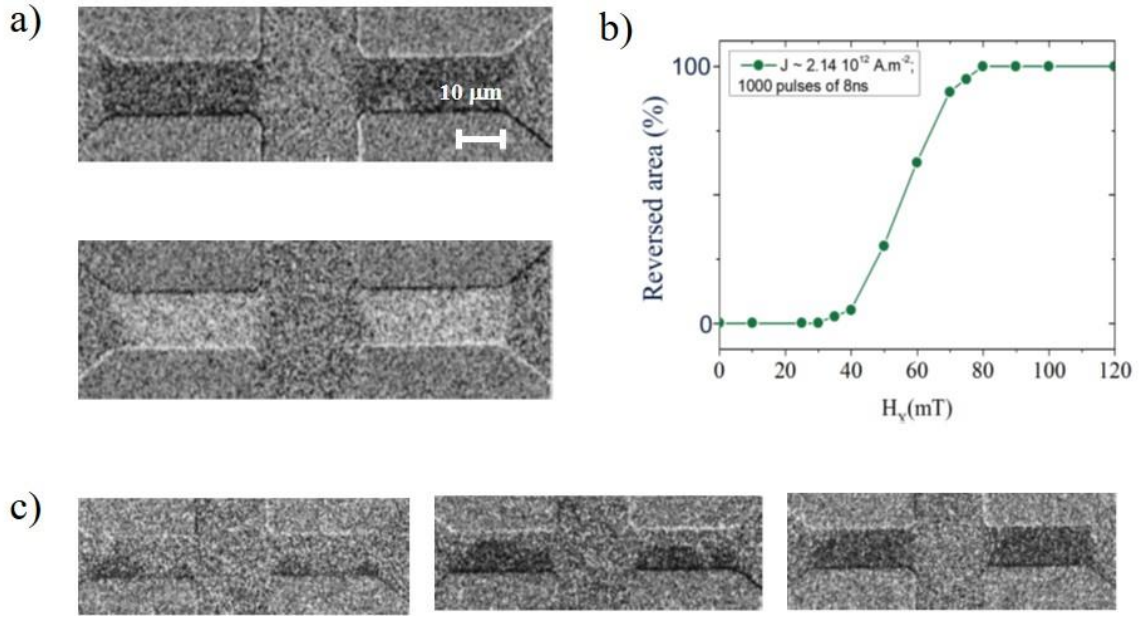

# References

1. Ioan Mihai Miron, Kevin Garello, Gilles Gaudin, Pierre-Jean Zermatten, Marius V. Costache, Stéphane Auffret, Sébastien Bandiera, Bernard Rodmacq, Alain Schuhl, and Pietro Gambardella, Nature **476**, 189 (2011).
2. Luqiao Liu, Chi-Feng Pai, Y. Li, H. W. Tseng, D. C. Ralph, and R. A. Buhrman, Science **336** (6081), 555 (2012).
3. Murat Cubukcu, Olivier Boulle, Marc Drouard, Kevin Garello, Can Onur Avci, Ioan Mihai Miron, Juergen Langer, Berthold Ocker, Pietro Gambardella, and Gilles Gaudin, Applied Physics Letters **104** (4), 042406 (2014).
4. Kevin Garello, Can Onur Avci, Ioan Mihai Miron, Manuel Baumgartner, Abhijit Ghosh, Stéphane Auffret, Olivier Boulle, Gilles Gaudin, and Pietro Gambardella, Applied Physics Letters **105** (21), 212402 (2014).
5. C. Zhang, S. Fukami, H. Sato, F. Matsukura, and H. Ohno, Applied Physics Letters **107** (1), 012401 (2015).
6. Junyeon Kim, Jaivardhan Sinha, Masamitsu Hayashi, Michihiko Yamanouchi, Shunsuke Fukami, Tetsuhiro Suzuki, Seiji Mitani, and Hideo Ohno, Nature Materials **12**, 240 (2012).
7. Chaoliang Zhang, Michihiko Yamanouchi, Hideo Sato, Shunsuke Fukami, Shoji Ikeda, Fumihiro Matsukura, and Hideo Ohno, Applied Physics Letters **103** (26), 262407 (2013).
8. Kevin Garello, Ioan Mihai Miron, Can Onur Avci, Frank Freimuth, Yuriy Mokrousov, Stefan Blügel, Stéphane Auffret, Olivier Boulle, Gilles Gaudin, and Pietro Gambardella, Nature Nanotechnology **8**, 587 (2013).
9. Can Onur Avci, Kevin Garello, Corneliu Nistor, Sylvie Godey, Belén Ballesteros, Aitor Mugarza, Alessandro Barla, Manuel Valvidares, Eric Pellegrin, Abhijit Ghosh, Ioan Mihai Miron, Olivier Boulle, Stephane Auffret, Gilles Gaudin, and Pietro Gambardella, Physical Review B **89** (21), 214419 (2014).
10. Can Onur Avci, Kevin Garello, Johannes Mendil, Abhijit Ghosh, Nicolas Blasakis, Mihai Gabureac, Morgan Trassin, Manfred Fiebig, and Pietro Gambardella, Applied Physics Letters **107** (19), 192405 (2015).
11. Chi-Feng Pai, Luqiao Liu, Y. Li, H. W. Tseng, D. C. Ralph, and R. A. Buhrman, Applied Physics Letters **101** (12), 122404 (2012).
12. Chi-Feng Pai, Minh-Hai Nguyen, Carina Belvin, Luis Henrique Vilela-Leão, D. C. Ralph, and R. A. Buhrman, Applied Physics Letters **104** (8), 082407 (2014).
13. Soonha Cho, Seung-heon Chris Baek, Kyeong-Dong Lee, Younghun Jo, and Byong-Guk Park, Scientific Reports **5**, 14668 (2015).
14. M. S. Gabor, Jr. T. Petrisor, R. B. Mos, A. Mesaros, M. Nasui, M. Belmeguenai, F. Zighem, and C. Tiusan, Journal of Physics D: Applied Physics **49** (36), 365003 (2016).
15. Kai-Uwe Demasius, Timothy Phung, Weifeng Zhang, Brian P. Hughes, See-Hun Yang, Andrew Kellock, Wei Han, Aakash Pushp, and Stuart S. P. Parkin, Nature Communications **7**, 10644 (2016).
16. C. Zhang, S. Fukami, K. Watanabe, A. Ohkawara, S. DuttaGupta, H. Sato, F. Matsukura, and H. Ohno, Applied Physics Letters **109** (19), 192405 (2016).
17. Luqiao Liu, Takahiro Moriyama, D. C. Ralph, and R. A. Buhrman, Physical Review Letters **106** (3), 036601 (2011).
18. Luqiao Liu, O. J. Lee, T. J. Gudmundsen, D. C. Ralph, and R. A. Buhrman, Physical Review Letters **109** (9), 096602 (2012).

19  Can Onur Avci, Kevin Garello, Ioan Mihai Miron, Gilles Gaudin, Stéphane Auffret, Olivier Boulle, and Pietro Gambardella, Applied Physics Letters **100** (21), 212404 (2012).
20  Hwang-Rae Lee, Kyujoon Lee, Jaehun Cho, Young-Ha Choi, Chun-Yeol You, Myung-Hwa Jung, Frédéric Bonell, Yoichi Shiota, Shinji Miwa, and Yoshishige Suzuki, Scientific Reports **4**, 6548 (2014).
21  Abhijit Ghosh, Kevin Garello, Can Onur Avci, Mihai Gabureac, and Pietro Gambardella, Physical Review Applied **7** (1), 014004 (2017).
22  Minh-Hai Nguyen, Chi-Feng Pai, Kayla X. Nguyen, David A. Muller, D. C. Ralph, and R. A. Buhrman, Applied Physics Letters **106** (22), 222402 (2015).
23  S. V. Aradhya, G. E. Rowlands, J. Oh, D. C. Ralph, and R. A. Buhrman, Nano Letters **16** (10), 5987 (2016).
24  M. Cubukcu, O. Boulle, N. Mikuszeit, C. Hamelin, T. Brächer, N. Lamard, M. Cyrille, L. Buda-Prejbeanu, K. Garello, I. M. Miron, O. Klein, G. de Loubens, V. V. Naletov, J. Langer, B. Ocker, P. Gambardella, and G. Gaudin, IEEE Transactions on Magnetics **54** (4), 1 (2018).
25  K. Garello, F. Yasin, S. Couet, L. Souriau, J. Swerts, S. Rao, S. Van Beek, W. Kim, E. Liu, S. Kundu, D. Tsvetanova, N. Jossart, K. Croes, E. Grimaldi, M. Baumgartner, D. Crotti, A. Furnémont, P. Gambardella, and G.S. Kar, arXiv:1810.10356 (2018).
26  S. Ikeda, K. Miura, H. Yamamoto, K. Mizunuma, H. D. Gan, M. Endo, S. Kanai, J. Hayakawa, F. Matsukura, and H. Ohno, Nature Materials **9**, 721 (2010).
27  B. Dieny and M. Chshiev, Reviews of Modern Physics **89** (2), 025008 (2017).
28  M. S. Gabor, T. Petrisor, C. Tiusan, and M. Hehn, Physical Review B **84** (13), 134413 (2011).
29  Zhenchao Wen, Hiroaki Sukegawa, Shinya Kasai, Masamitsu Hayashi, Seiji Mitani, and Koichiro Inomata, Applied Physics Express **5** (6), 063003 (2012).
30  Can Onur Avci, Kevin Garello, Mihai Gabureac, Abhijit Ghosh, Andreas Fuhrer, Santos F. Alvarado, and Pietro Gambardella, Physical Review B **90** (22), 224427 (2014).
31  Minh-Hai Nguyen, D. C Ralph, and R. A Buhrman, Physical Review Letters **116** (12), 126601 (2016).
32  Edmund Clifton Stoner and E. P. Wohlfarth, Philosophical Transactions of the Royal Society of London. Series A, Mathematical and Physical Sciences **240** (826), 599 (1948).
33  J.-C. Rojas-Sánchez, P. Laczkowski, J. Sampaio, S. Collin, K. Bouzehouane, N. Reyren, H. Jaffrès, A. Mougin, and J.-M. George, Applied Physics Letters **108** (8), 082406 (2016).
34  Manuel Baumgartner, Kevin Garello, Johannes Mendil, Can Onur Avci, Eva Grimaldi, Christoph Murer, Junxiao Feng, Mihai Gabureac, Christian Stamm, Yves Acremann, Simone Finizio, Sebastian Wintz, Jörg Raabe, and Pietro Gambardella, Nature Nanotechnology **12**, 980 (2017).